\documentclass[twocolumn, prb,superscriptaddress]{revtex4-1}
\usepackage{amsmath}
\usepackage{float}
\usepackage{epstopdf}
\usepackage{verbatim}
\usepackage{amsfonts,amssymb}
\usepackage{graphicx}
\usepackage{hyperref}    % links
\usepackage{bm}          % bold math letters
\usepackage{natbib}
\usepackage{soul} % in order to highlight
\usepackage{color}
\usepackage{longtable}
\usepackage{ textcomp }

\begin{document}

\title{Collapse of electrons to a donor cluster in SrTiO$_3$}

\author{Han Fu}
\affiliation{Fine Theoretical Physics Institute, University of Minnesota, Minneapolis, MN 55455, USA}
\author{K. V. Reich}
\email{Reich@mail.ioffe.ru}
\affiliation{Fine Theoretical Physics Institute, University of Minnesota, Minneapolis, MN 55455, USA}
\affiliation{Ioffe Institute, St Petersburg, 194021, Russia}
\author{B. I. Shklovskii}
\affiliation{Fine Theoretical Physics Institute, University of Minnesota, Minneapolis, MN 55455, USA}

\begin{abstract}
It is known that a nucleus with charge $Ze$, where $Z>170$ creates electron-positron pairs from the vacuum. Electrons collapse onto the nucleus resulting in a net charge $Z_n < Z$ while the positrons are emitted. This effect is due to the relativistic dispersion law. The same reason leads to the collapse of electrons to the donor cluster with a large charge number $Z$ in narrow-band gap semiconductors, Weyl semimetals and graphene. In this paper, a similar effect of electron collapse and charge renormalization is found for a donor cluster in SrTiO$_3$ (STO), but with a different origin. At low temperatures, STO has an enormously large dielectric constant and the nonlinear dielectric response becomes dominant when the electric field is still small. This leads to the collapse of electrons into a charged spherical donor cluster with radius $R$ when its total charge number $Z$ exceeds a critical value $Z_c\simeq R/a$ where $a$ is the lattice constant. The net charge $Z_{n}e$ grows with $Z$ until $Z$ exceeds $Z^*\simeq(R/a)^{9/7}$. After this point, the charge of the compact core $Z_{n}$ remains $\simeq Z^*$, while the rest $Z^*$ electrons form a sparse Thomas-Fermi electron atmosphere around it. We show that the thermal ionization of such two-scale atoms easily strips the outer atmosphere while the inner core remains preserved. We extend our results to the case of long cylindrical clusters. We discuss how our predictions can be tested by measuring conductivity of chain of discs of charge on the STO surface.
\end{abstract}

\date{\today}

\maketitle

\section{Introduction}

Recently, studies of ABO$_3$ perovskite crystals have been a subject of interest due to their intriguing magnetic, superconducting, and multiferroic properties \cite{Rondinelli} and subsequent significant technological applications. Among them, SrTiO$_3$ (STO) has attracted special attention \cite{Stemmer, Zubko}. STO is a semiconductor with a band gap of $\simeq$ 3.2 eV and a large dielectric constant $\kappa=2\times10^4 $ at liquid helium temperature. Like the conventional semiconductors, STO can be used as the basis for a number of devices \cite{Hwang, Xie}.

Many of the devices are realized by doping the bulk STO such as the reduction of STO through generating oxygen vacancies at high temperatures. The vacancies either form along a network of extended defects\cite{Ov2002} or assemble together to lower the system's energy\cite{Ov2007, Muller_2004}, probably producing large positively charged donor clusters. Another way to more controllably create such a cluster is to ``draw" a disc of charge by the atomic force microscope (AFM) tip on the surface of LAO/STO structure with the subcritical thickness for LaAlO$_3$ (LAO) \cite{Cen_2008,*Cen_2010}. The potential caused by such a positive disc in the bulk STO is similar to that of a charged sphere.

Let's consider a spherical donor cluster with radius $R$ and charge $Ze$. We assume that the cluster is located on the background of uniformly n-type doped STO in which the Fermi level is very close to the bottom of the conduction band. There are $Z$ electrons located at distances from $\kappa b/Z$ to the Bohr radius $\kappa b$ from the cluster, which form a Thomas-Fermi ``atom" \cite{Landau} with it. Here $b=\hbar^2/m^*e^2$, $m^*\approx1.8m_e$ is the effective electron mass in STO \cite{Ahrens} with $m_e$ being the electron mass. Since $\kappa$ is large, the electrons are far away from the cluster and the whole ``atom" is very big. As $Z$ increases, the electron gas swells inward to hold more electrons. However, we find that when $Z$ goes beyond a certain value $Z_c$ ($\kappa b/Z$ is still much larger than $R$ at this moment), the physical picture is qualitatively altered. Surrounding electrons start to collapse into the cluster and the net cluster charge gets renormalized from $Ze$ to $Z_ne$ with $Z_n\ll Z$ at very large $Z$.

The effect of charge renormalization is not new \cite{Pomeranchuk, *Zeldovich, Kolomeisky}. For a highly charged nucleus with charge $Ze$, the vacuum is predicted to be unstable against creation of electron-positron pairs, resulting in a collapse of electrons onto the nucleus with positrons emitted \cite{Pomeranchuk,Zeldovich}. This instability happens when $Z>Z_c$ with $Z_c\simeq170\gtrsim1/\alpha$, where $\alpha=e^2/\hbar c\simeq1/137$ is the fine structure constant. When $Z$ exceeds $Z^*\simeq1/\alpha^{3/2}\simeq137^{3/2}$, the net charge saturates at $Z^*$ \cite{Kolomeisky}. In the condensed matter setting, there are similar phenomena in narrow-band gap semiconductors and Weyl semimetals \cite{Kolomeisky} as well as graphene \cite{Fogler, *Novikov, *Levitov,*Pereira, *Gorsky, * Crommie}.  In all these cases, the collapse happens because the energy dispersion of electrons is relativistic in the Coulomb field of a compact donor cluster playing the role of a nucleus. In our work, however, the collapse originates from the strong nonlinearity of dielectric constant in STO at small distances from the cluster. In the case of a spherical donor cluster, this nonlinearity leads to the change of the attractive potential near the cluster from being $\propto 1/r$ to $\propto1/r^5$, resulting in the collapse of electrons to the cluster.

\begin{figure}[h]
$\begin{array}{c}
\includegraphics[width=0.47\textwidth]{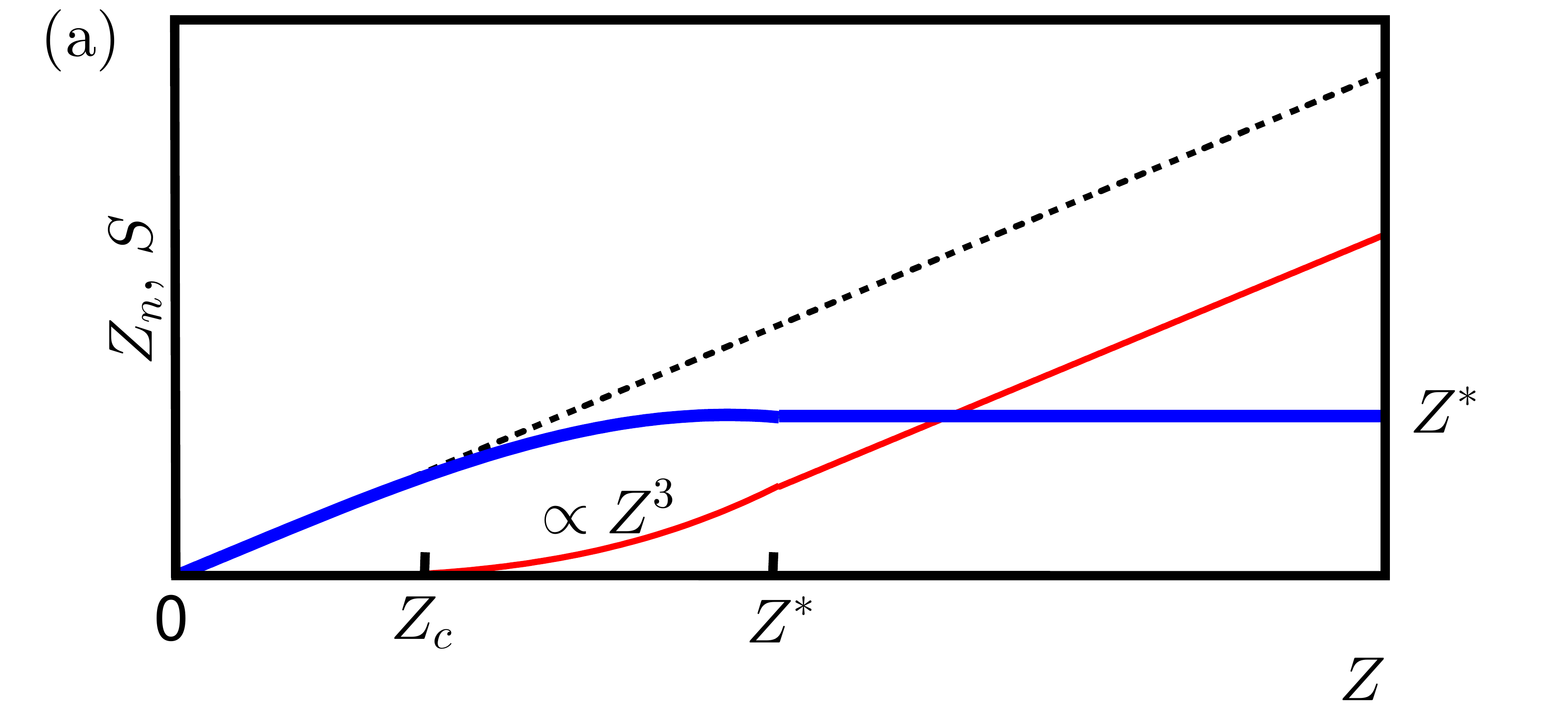}\\
\includegraphics[width=0.47\textwidth]{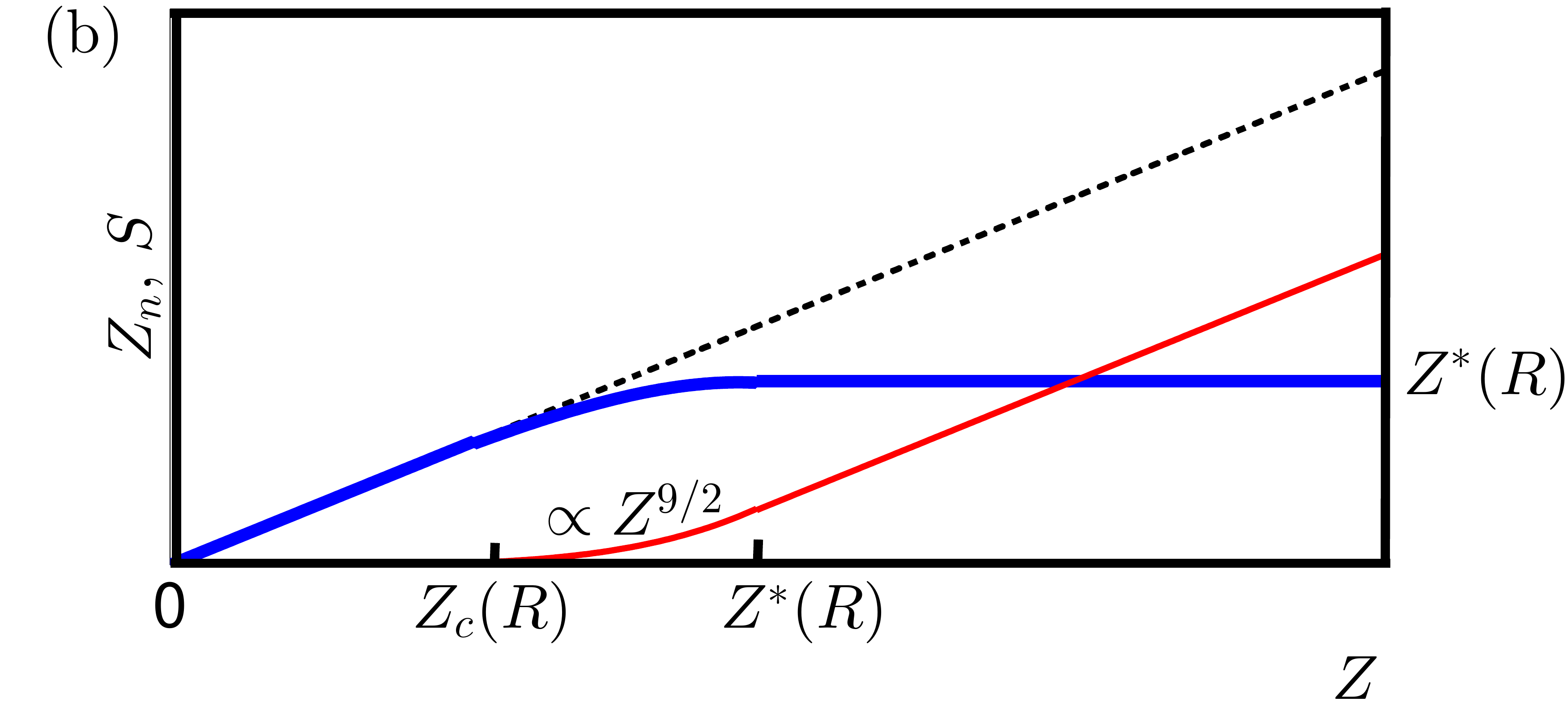}
\end{array}$
\caption{(Color online) The number of collapsed electrons $S$ and the renormalized net charge $Z_ne$ as a function of the original charge $Ze$. $S$ is shown by the thin solid line (red), $Z_n$ is denoted by the thick solid line (blue), and the dashed line (black) is a guide-to-eye where $Z_n=Z$. $Z_c$ denotes the critical value where electrons begin to collapse and $Z^*$ is the saturation point where $Z_n$ stops growing. (a) Collapse of electrons and charge renormalization for highly charged nuclei. $S\propto Z^3$ at $Z_c\ll Z\ll Z^*$ \cite{Migdal_1977}. (b) Collapse of electrons and charge renormalization for spherical donor clusters in STO. $S\propto Z^{9/2}$ at $Z_c\ll Z\ll Z^*$.}\label{fig:1}
\end{figure}

The phenomena of electron collapse and charge renormalization in both heavy nuclei and our work are presented in Fig. \ref{fig:1}. In our case, the first electron collapses at $Z\simeq Z_c\simeq R/a$, where $a$ is the lattice constant, and at $Z\gg Z^*\simeq (R/a)^{9/7}$, the net charge of nucleus $Z_ne$ saturates as $Z_n\simeq Z^*$.

In the remainder of this paper, we use the Thomas-Fermi approximation to show how the electron gas collapses into the cluster at $Z\gg Z_c$ and find the corresponding electron density. In Sec. II, we demonstrate that when $Z_c\ll Z\ll Z^*$, the charge renormalization is relatively weak and the final charge number $Z_n$ is just a little bit below $Z$. On the other hand, when $Z\gg Z^*$, the renormalization is very strong and $Z_n$ is maintained at the level of $Z^*$. The inner core of the cluster atom with charge $Z^*$ is surrounded by a sparse Thomas-Fermi atmosphere of $Z^*$ electrons in which the large linear dielectric constant plays the main role. In Sec. III, we generalize our studies to the cylindrical donor clusters and introduce the notion of maximum linear charge density $\eta*$ similar to $Z^*$ which is got when the bare charge density of the cluster  $\eta$ is large. In Sec. IV we consider the thermal ionization of donor cluster atoms and arrive at the conclusion that for both spheres and cylinders the external electron atmosphere is easily ionized while the inner core with charge $Z^*$ or linear charge density $\eta*$ is robust against the ionization. We suggest experimental verification of our theory for a periodic chain of charge discs created by methods of Ref. \onlinecite{Cen_2008,*Cen_2010}. If neighboring disc atoms overlap via the outer electron atmosphere (the ``bridges"), raising temperature may ionize these bridges and sharply reduce the chain conductance. We conclude our work in Sec. V.
%\pacs{73.20.-r,71.10.Ca,73.30.+y,68.47.Gh}
%73.20.-r Electron states at surfaces and interfaces
%71.10.Ca	Electron gas, Fermi gas
%73.30.+y Surface double layers
%68.47.Gh	Oxide surfaces

\section{Spherical Donor Clusters}
\subsection{Nonlinear Dielectric Response}
It is well known that STO is a quantum paraelectric where the onset of ferroelectric order is suppressed by quantum fluctuations \cite{Nakamura}. In such ferroelectric-like materials, the free energy density $F$ can be expressed by the Landau-Ginzburg theory as a power-series expansion of the polarization $P$ \cite{Ginzburg}:
\begin{equation}
F=F_0+\frac{\tau}{2}P^2+\frac{A}{4P_0^2}P^4-EP
\nonumber
\end{equation}
(in Gaussian units), where $F_0$ stands for the free energy density at $P=0$, $E$ is the electric field, $\tau=4\pi/(\kappa-1)\simeq 4\pi/\kappa$ is the inverse susceptibility, $\kappa\gg1$ is the dielectric constant, $A\approx0.9$ \cite{RS}, $P_0=e/a^2$, $a=3.9\,\AA$ is the lattice constant. We neglect the gradient terms of polarization since they play a minor role in the nonlinear regime. We assume that the dielectric response is isotropic. This is justified by the small anisotropy in the nonlinear response of STO\cite{Fleury}. The crystal polarization $P$ is determined by minimizing the free energy density $F$ in the presence of the electric field $E$, i.e., $\delta F/\delta P = 0$. This gives
\begin{equation}
E=\frac{4\pi P}{\kappa}+\frac{A P^3}{P_0^2}.
\nonumber
\end{equation}
For very big $\kappa$, the nonlinear term in this expression is likely to dominate over the linear one even at relatively small $P$ as long as $P\gg\sqrt{4\pi/\kappa A}P_0$. At distances not too far from the cluster, this relationship can easily be satisfied and the dielectric response becomes nonlinear \cite{RS}:
\begin{equation}
E=\frac{A P^3}{P_0^2}.
\label{eq:1}
\end{equation}
Since $P_0=e/a^2$ is really big, we can expect that $P<\sqrt{4\pi/A}P_0$ always holds, which gives $E<4\pi P$. Then,
\begin{equation}
D=E+4\pi P\approx 4\pi P.
\label{eq:2}
\end{equation}
According to the Gauss's law in dielectric media, we know $\nabla\cdot D=4\pi\rho$ where $\rho$ is the free charge density. Together with Eqs. (\ref{eq:1}) and (\ref{eq:2}), we have
\begin{equation}
\nabla\cdot (\nabla\phi)^{1/3}=-\left(\frac{A}{P_0^2}\right)^{1/3}\rho
\label{eq:x}
\end{equation}
where $\phi$ is the electric potential.

Below, we show that this nonlinear dielectric response leads to the collapse of electrons into the spherical donor cluster inside STO when the cluster charge is large enough.

\subsection{Renormalization of Charge}

\begin{figure}[h]
\includegraphics[width=0.5\textwidth]{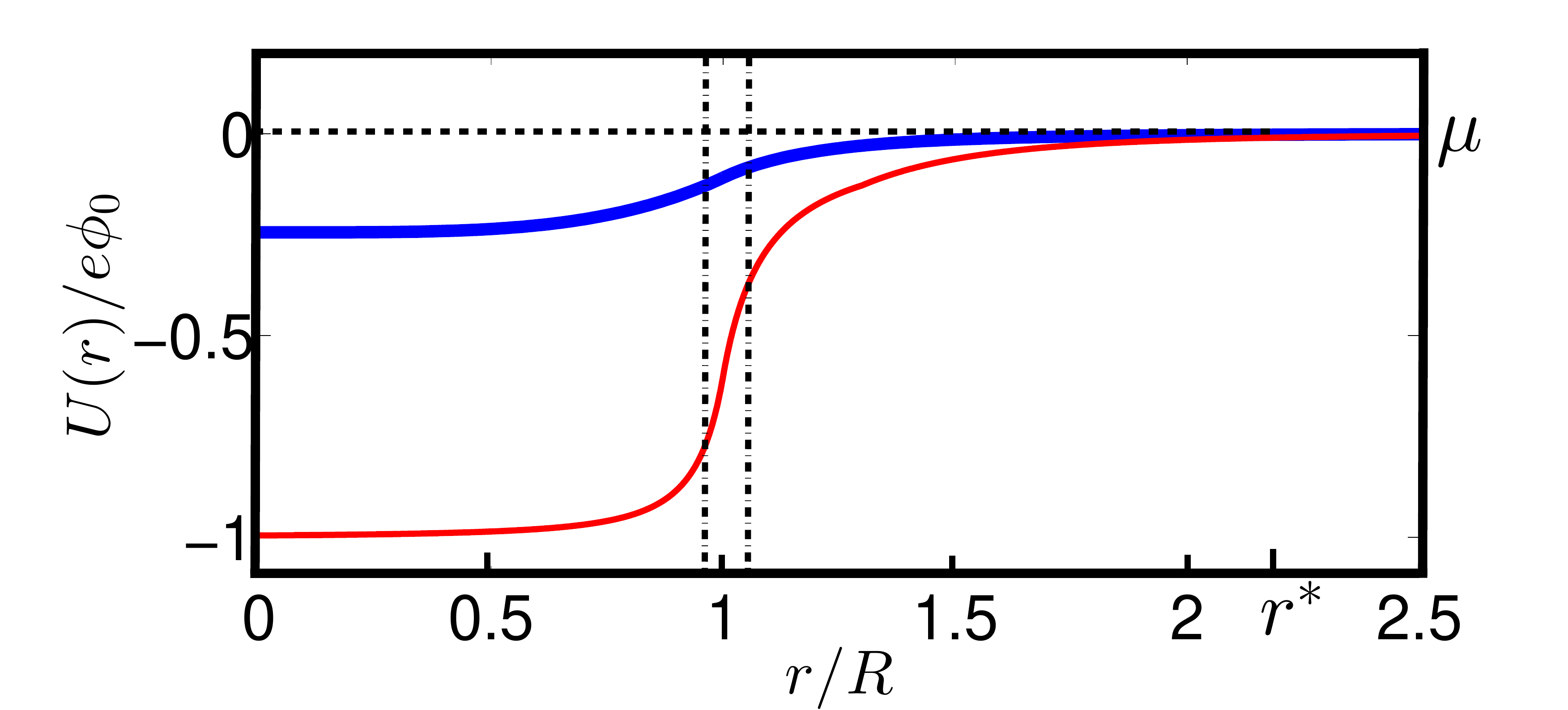}
\caption{(Color online) Potential energy of electrons $U(r)=-e\phi(r)$ as a function of radius $r$. $\phi_0$ is defined as $n(\phi_0)=n_0$, where $n_0=3Z/4\pi R^3$, $n(r)$ is a function of $\phi(r)$ given by Eq. (\ref{eq:5}). The thick solid line (blue) represents the potential profile of a cluster of charge $Z\lesssim Z^*$ which is in the regime of weak charge renormalization. The thin solid line (red) represents the potential of a cluster at $Z\gg Z^*$ in the strong renormalization regime, where the two vertical dotted lines show edges of the ``double-layer" structure of width $\sim d\ll R$. The horizontal dashed line (black) indicates the position of the chemical potential $\mu= 0$. $r^*$ is the external radius of the collapsed electron gas where Thomas-Fermi approach fails.}
\label{fig:3}
\end{figure}

Consider a large spherical donor cluster of the radius $R$ and the total positive charge $Ze$ such that $a\ll R< \kappa b/Z$. If the dielectric response is linear, the electrons are mainly located at distances between $r_1=\kappa b/Z$ and $r_A=\kappa b$ from the cluster \cite{Landau} where $b=\hbar^2/m^*e^2\approx0.29\,\AA$ \cite{RS}. For a very large $\kappa$, these radii are huge ($r_A=700$ nm in STO at liquid helium temperature where $\kappa=20000$) and the electrons are far away from the cluster. However, at small distances, the dielectric response is nonlinear and changes the potential form. If the potential energy outweighs the kinetic energy, electrons are attracted to the cluster and renormalize the net charge. To see when this will happen, we look at the specific form of electric potential in this situation. We can calculate the potential from Eq. (\ref{eq:x}). But due to the simple charge distribution here, we can get it in an easier way.  At $r>R$ where $r$ is the distance from the cluster center, the sphere looks like a point charge and $D(r)=Ze/r^2$. Using this together with Eqs. (\ref{eq:1}) and (\ref{eq:2}), one can calculate the electric field \begin{comment}$E(r)$:
\begin{subequations}
\begin{align}
E(r)&=\frac{A}{P_0^2}\left(\frac{Ze}{4\pi}\right)^3\frac{r^3}{R^9}, 0<r<R\\
E(r)&=\frac{A}{P_0^2}\left(\frac{Ze}{4\pi}\right)^3\frac{1}{r^6},\,R<r\ll r_1.
\end{align}
\end{subequations}
Correspondingly,\end{comment}
and get the electric potential $\phi(r)$ as:
\begin{equation}
\phi(r)=\frac{A}{P_0^2}\left(\frac{Ze}{4\pi}\right)^3\frac{1}{5r^5}, \quad\quad\quad\quad\quad\quad R<r\ll r_1
\label{eq:3b}
\end{equation}
with $\phi(r=\infty)$ defined as zero. Inside the cluster at $r<R$, since the charge is uniformly distributed over the sphere, the total positive charge enclosed in the sphere of radius $r$ is equal to $Zer^3/R^3$, so $D(r)=Zer/R^3$. One then gets the corresponding potential $\phi(r)$:
\begin{equation}
\phi(r)=\frac{A}{P_0^2}\left(\frac{Ze}{4\pi}\right)^3\left(\frac{9}{20}\frac{1}{R^5}-\frac{1}{4}\frac{r^4}{R^9}\right), 0<r<R
\label{eq:3a}
\end{equation}
using the boundary condition $\phi(r=R^-)=\phi(r=R^+).$
A schematic graph of the potential energy $U(r)=-e\phi(r)$ is shown in Fig. \ref{fig:3} by the thick solid line (blue).

The Hamiltonian for a single electron is $H=p^2/2m^*-e\phi(r)$, where $p$ is the momentum of the electron and $m^*$ is the effective electron mass in STO \cite{RS}. If we approximately set $p\simeq\hbar/2r$, we get a positive total energy of the electron everywhere when $Z$ is very small. This means there are no bound states of electron in the cluster. However, when $Z$ is big enough so that $Z>Z_c$, the electron can have negative total energy at $r<R$ and will collapse into the cluster. \begin{comment}$Z_c$ is found when the zero energy position first appears in space at $r=r_0<R$:
\begin{equation}
\begin{aligned}
\left(\frac{\hbar}{2r_0}\right)^2\frac{1}{2m_e}-e\phi(r_0)\quad \quad &=0\\
%\frac{eA}{P_0^2}\left(\frac{Z_ce}{4\pi}\right)^3\left(\frac{9}{20}\frac{1}{R^5}-\frac{1}{4}\frac{r_0^4}{R^9}\right)&=0\\
\frac{d}{dr}\left[\left(\frac{\hbar}{2r}\right)^2\frac{1}{2m_e
}-e\phi(r)\right]_{r=r_0}&=0
%\frac{eA}{P_0^2}\left(\frac{Z_ce}{4\pi}\right)^3\left(\frac{9}{20}\frac{1}{R^5}-\frac{1}{4}\frac{r^4}{R^9}\right)\right]_{r=r_0}&=0.
\end{aligned}
\nonumber
\end{equation}
So,\end{comment}
Using Eqs. (\ref{eq:3b}) and (\ref{eq:3a}), we find
\begin{equation}
%r_0&=0.88R\\%\left(\frac{3}{5}\right)^{1/4}\\
Z_c\approx\frac{4\pi(b/Aa)^{1/3}R}{a}% \left[\frac{1}{4\left(3/5\right)^{3/2}}\right]^{1/3}
\sim \frac{R}{a}\gg 1.
\end{equation}
As $Z$ continues increasing, more and more electrons get inside the cluster filling it from the center where the potential energy is lowest (see Fig. \ref{fig:3}). The single-electron picture no longer applies. Instead, we use the Thomas-Fermi approximation \cite{Landau} with the electrochemical potential $\mu=0$, which gives:
\begin{equation}
n(r)=\frac{c_1}{b^3}\left[\frac{\phi(r)}{e/b}\right]^{3/2},
\label{eq:5}
\end{equation}
where $n(r)$ is the electron density at radius $r$, $c_1=2^{3/2}/3\pi^2\approx 0.1$.

We assume that the bulk STO is a heavily doped semiconductor in which the Fermi level lies in the conduction band. On the other hand, due to the relatively high effective electron mass, the Fermi energy is much smaller than the depth of the potential well shown in Fig. \ref{fig:3} and is thus ignored. Since at $r=\infty$ the electric potential $\phi(r)$ is defined as 0, we then have the electrochemical potential $\mu\simeq 0$.

When the number of collapsed electrons $S$ is small, their influence on the electric potential is weak. One can still use Eqs. (\ref{eq:3b}) and (\ref{eq:3a}) for $\phi(r)$ and get the corresponding expression of $n(r)$. At $r>R$, since $\phi(r)$ is $\propto 1/r^5$, we get $n(r)\propto1/r^{15/2}$. In this way, we calculate $S$ as
\begin{equation}
S=\int^\infty_0 n(r)4\pi r^2dr=0.5Z\left(\frac{Z}{Z^*}\right)^{7/2}\propto Z^{9/2},
\label{eq:6}
\end{equation}
where %the electrons collapsed onto the surface of the cluster instead of being inside accounts for $0.1(Z/Z^*)^{7/2}Z$ and
\begin{equation}
Z^*=\left[\frac{4\pi(b/Aa)^{1/3}R}{a}\right]^{9/7},
\label{eq:7}
\end{equation}
The net charge number of the cluster is
\begin{equation}
Z_n=Z-S=Z\left[1-0.5\left(\frac{Z}{Z^*}\right)^{7/2}\right].
\label{eq:8}
\end{equation}
One can see, when $Z_c\ll Z\ll Z^*$, one gets $S\ll Z$ and $Z_n\lesssim Z$, meaning the charge renormalization is weak. However, at $Z\sim Z^*$, according to Eqs. (\ref{eq:6}) and (\ref{eq:8}), we get $Z_n\sim S\sim Z^*$. The potential contributed by electrons is no longer perturbative. This brings us to the new regime of strong renormalization of charge.

$\phantom{}$

We show that at $Z\gg Z^*$ the net charge $Z_ne$ saturates at the level of $Z^*e$. Indeed, when $Z$ grows beyond $Z^*$, $Z_n$ can not go down and therefore can't be much smaller than $Z^*$.
At the same time it can not continue going up, otherwise as follows from Eqs. (\ref{eq:3b}) and (\ref{eq:6}) with $Z$ replaced by $Z_n\gg Z^*$, the total electron charge surrounding the charge $Z_ne$ at $r > R$ would become $Se \simeq Z_ne (Z_n/Z^*)^{7/2}\gg Z_ne$ leading to a negative charge seen from infinity. Thus, at $Z\gg Z^*$, the net charge $Z_n$ saturates at the universal value of the order of $Z^*$ as is shown  in Fig. \ref{fig:1}b. As we emphasized in Fig. \ref{fig:1}, this result is qualitatively similar to the one obtained for heavy nuclei and donor clusters in Weyl semimetals and narrow-band gap semiconductors in Ref. \onlinecite{Kolomeisky}.

In the following subsection, we show how the renormalization of charge at $Z\gg Z^*$ is realized through certain distribution of electrons, in which a structure of ``double layer" (see Fig. \ref{fig:3}) plays an important role.
\subsection{Radial Distribution of Electrons}

\begin{comment}At $Z_c<Z<Z^*$, a small quantity of electrons collapse to the cluster, most of which are located at $r<R$. Outside the cluster sphere, since the potential decays rapidly as $1/r^5$, the radial concentration of electrons $n(r)r^2$ also diminishes fast as $1/r^{11/2}$:
\begin{equation}
n(r)r^2=c_2Z\left(\frac{Z}{Z^*}\right)^{7/2}\frac{1}{a^3}\left(\frac{R}{a}\right)^{9/2}\left(\frac{a}{r}\right)^{11/2},
\label{eq:9}
\end{equation}
where $c_2\approx 0.04$.
\end{comment}

At $Z\gg Z^*$, the charge renormalization is strong and the most of the sphere of radius $R$ is completely neutralized by electrons. In the neutral center of the sphere, the electron density $n(r)=n_0$, where $n_0=3Z/4\pi R^3$ is the density of the positive charge inside the cluster. The corresponding ``internal" electric potential $\phi_{in}(r)=\phi_0$ where $\phi_0$ is given by $n(\phi_0)=n_0$ using Eq. (\ref{eq:5}). $\phi_{in}(r)$ is then $\propto (n_0a^3)^{2/3}\propto [Z/(R/a)^3]^{2/3}$. Outside the cluster, when the charge is renormalized to $Z_n$, one gets a potential $\phi_{out}(r)$ similar to Eq. (\ref{eq:3b}) with $Z$ replaced by $Z_n$. Since $Z_n$ is $\sim Z^*$ where $Z^*$ is given by Eq. (\ref{eq:7}), we get $\phi_{out}(r)\propto (R/a)^{-8/7}$ at a distance $r$ of the order $R$. Thus, close to the cluster surface, the ratio of the outside potential $\phi_{out}(r)$ to the inside potential $\phi_{in}(r)$ is $\simeq(R/a)^{6/7}/Z^{2/3}\ll 1 $ since $Z\gg Z^*\simeq(R/a)^{9/7}$. This indicates a sharp potential drop across the sphere surface.

At $0<R-r\ll R$, there's a thin layer of uncompensated positive charges. At $0<r-R\ll R$, a higher potential than farther away means a larger electron concentration that forms a negative layer close to the surface. This ``double-layer" structure resembles a capacitor which quickly brings the potential down across the surface as shown in Fig. \ref{fig:3}. An analogous structure also exists in heavy nuclei \cite{Pomeranchuk, Migdal_1977} with charge $Z\gg1/\alpha^{3/2}$.

To make the analysis more quantitative, one needs to know the specific potential profile in this region. Using Eq. (\ref{eq:x}), we get the general equation of $\phi(r)$ in the spherical coordinate system:
\begin{subequations}
\begin{align}
\left(\frac{d}{dr}+\frac{2}{r}\right)\left(\frac{d\phi}{dr}\right)^{1/3}&=\frac{A^{1/3}e}{P_0^{2/3}}\left[n(r)-n_0\right],\, r<R\label{eq:9a}\\
\left(\frac{d}{dr}+\frac{2}{r}\right)\left(\frac{d\phi}{dr}\right)^{1/3}&=\frac{A^{1/3}e}{P_0^{2/3}}n(r), \quad\quad\,\quad r>R\label{eq:9b}
\end{align}
\end{subequations}
Near the cluster surface, we can approximately use a plane solution of $\phi(r)$, i.e., ignore the $2/r$ term on the left side. This kind of solution for $r\gtrsim R$ is already known \cite{RS}:
\begin{equation}
\phi(r)=\frac{c_3}{A^{2/7}}\frac{e}{b}\left(\frac{b}{a}\right)^{16/7}\left(\frac{a}{x+d}\right)^{8/7},
\label{eq:10}
\end{equation}
where $x=r-R\ll R$ is the distance to the surface and $d\ll R$ is the characteristic decay length to be determined, $c_3\approx 6$. Correspondingly, the radial electron concentration at $r\gtrsim R$ is given by
\begin{equation}
\begin{aligned}
n(r)r^2&=\frac{c_4}{A^{3/7}}\frac{1}{b^3}\left(\frac{b}{a}\right)^{24/7}\left(\frac{a}{x+d}\right)^{12/7}r^2\\
&\approx\frac{c_4}{A^{3/7}}\frac{1}{b^3}\left(\frac{b}{a}\right)^{24/7}\left(\frac{a}{x+d}\right)^{12/7}R^2,
\end{aligned}
\label{eq:12}
\end{equation}
where $r\approx R$, $c_4\approx1$.

Since the ``double-layer" structure resembles a plane capacitor, near the surface, the potential drop is nearly linear with the radius. Using Eq. (\ref{eq:10}), one can get $\phi(r)\approx (1-8x/7d)\phi(R)$ at $0<x=r-R\ll d$, which gives the electric field $8\phi(R)/7d$ inside the ``double layer". At $r<R$, this electric field persists and gives $\phi(r)\approx (1+8x/7d)\phi(R)$ at $0<x=R-r\ll d$. As $r$ further decreases, the positive layer ends and the potential crosses over to the constant value $\phi_0$  given by $n(\phi_0)=n_0$ using Eq. (\ref{eq:5}). This boundary condition gives
\begin{equation}
d=\frac{c_5}{A^{1/4}}\left(\frac{b}{a}\right)^{1/4}\frac{a}{(n_0a^3)^{7/12}},
\label{eq:13}
\end{equation}
where $c_5\approx 2$.
By expressing $n_0$ in terms of $Z$ and $R$, we get $d/R\propto\left(Z^*/Z\right)^{7/12}\ll 1$ at $Z\gg Z^*$.

According to Eq. (\ref{eq:10}), when $x$ is comparable to $R$ and the plane approximation is about to lose its validity, $\phi(r)$ is $\propto(R/a)^{-8/7}$. It is weak enough to match the low electric potential $\phi_{out}(r)\propto(R/a)^{-8/7}$ caused by the renormalized charge $Z_n\sim Z^*$ at $r\sim R$. The plane solution then crosses over to the potential $\phi_{out}(r)\propto {Z^*}^3/r^5$ which is the asymptotic form at large distances.

\begin{figure}[h]
\includegraphics[width=0.5\textwidth]{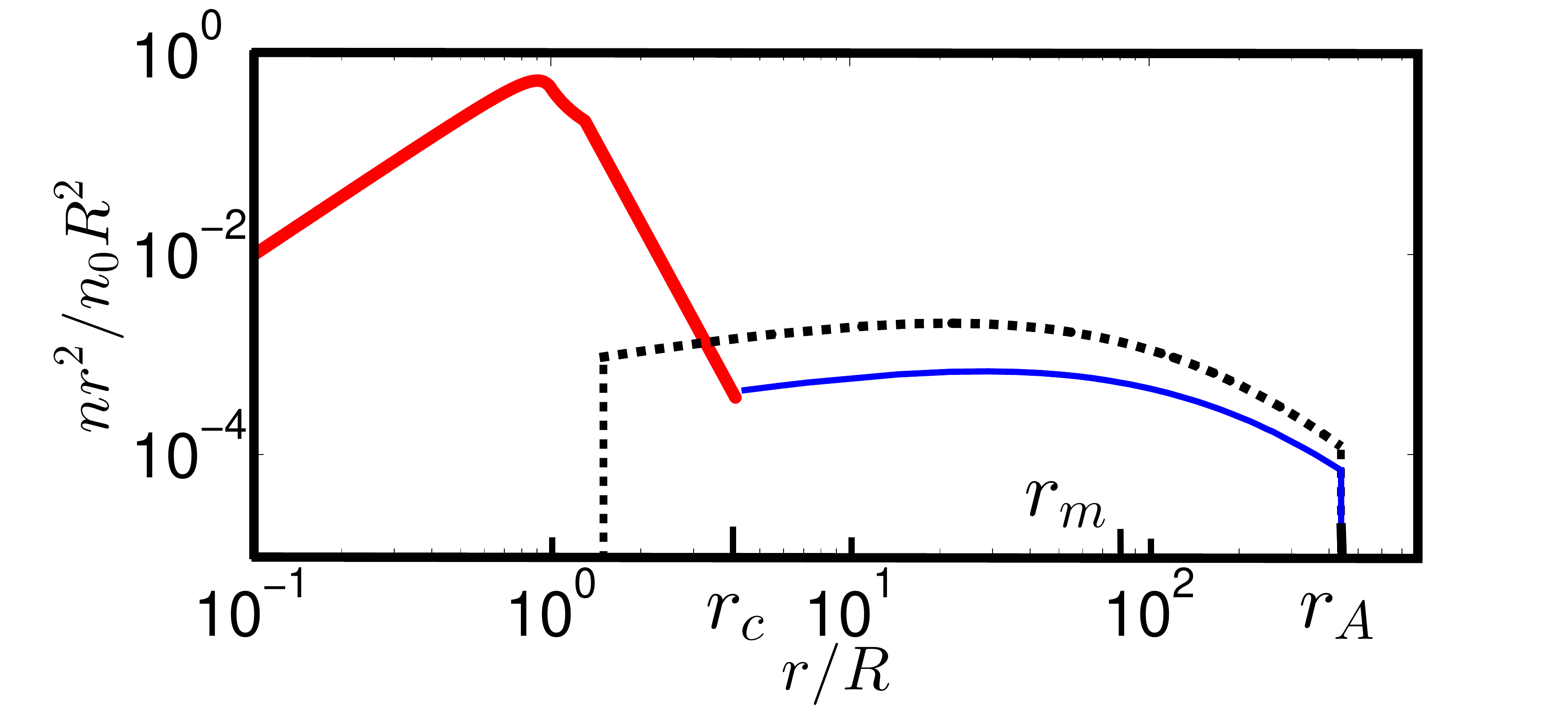}\\
\caption{(Color online) Radial electron concentration $n(r)r^2$ as a function of radius $r$. The thick solid line (red) represents the inner collapsed electrons at $r<r_c$ where the dielectric response is nonlinear. The thin solid line (blue) shows the electrons belonging to the outer shell which form the standard Thomas-Fermi atom with the renormalized nucleus of charge $Z^*$ at $r>r_c$, where the dielectric response is linear. This electron gas ends at the Bohr radius $r_A=\kappa b$ while most of them are at radius $r_m=\kappa b/{Z^*}^{1/3}$. The dashed line (black) denotes the electrons forming a Thomas-Fermi atom \cite{Landau} with a nucleus of charge $Z$ when $P_0$ is infinity and there's no range with nonlinear dielectric response. The reduction of electron density in the outer shell of electrons due to the collapse is substantial. The reason this is not immediately seen from the difference of height between the dashed line (black) and the thin solid line (blue) is that we use a logarithmic scale here. $n_0$ is defined as $3Z/4\pi R^3$. This graph is plotted at $b=0.35\,\AA$, $a=3.9\,\AA$, $A=0.9$, $R=4.4a$, $\kappa=20000$, $n_0=0.8/a^3$.}\label{fig:3a}
\end{figure}

A schematic plot of the potential energy $U(r)=-e\phi(r)$ as a function of radius $r$ is shown in Fig. \ref{fig:3} by the thin solid line (red). The corresponding radial distribution of electrons is shown in Fig. \ref{fig:3a} by the thick solid line (red).

$\phantom{}$

So far, we have got a $1/r^5$ potential $\phi(r)$ and $1/r^{11/2}$ radial electron concentration $n(r)r^2$ at $r\gg R$ in both weak and strong charge renormalization cases. However, as the electron density decreases to certain extent so that the Fermi wavelength $\lambda$ is comparable to the radius $r$, the gas is no longer degenerate and the Thomas-Fermi approach fails. Since $\lambda\simeq n(r)^{-1/3}$, we get this radius $r^*\simeq Z^*a$ at $Z\gg Z^*$. One should then return to the Schrodinger equation used for a single electron. Since the uncertainty principle estimates that the kinetic energy decays as $1/r^2$ while the potential energy is $\propto -{Z^*}^3/r^5$, the potential energy is smaller than the kinetic energy in magnitude at $r>r^*$, which means electrons can not stay at radii larger than $r^*$. One can also find that using the Thomas-Fermi solution $\phi_{out}(r)$, the total electron number calculated at $r>r^*$ is $\sim 1$, which again indicates there's no electron at $r>r^*$ considering the discreteness of electron charge. As a result, the $1/r^{11/2}$ tail of radial electron concentration will not continue to infinity but stop at radius $r^*$. This is a semi-classical result. Quantum mechanical analysis shows that the electron density does not go to zero right at $r^*$ but decays exponentially after this point. Since this decay is fast and brings very small corrections to the end of the inner electron gas, we don't consider it here.

At $\kappa=\infty$, the rest of the electrons are at the infinity so that we are dealing with a positive ion with charge $Z^*$. At finite but very large $\kappa$, at certain distance from the cluster, the field is so small that $P>\sqrt{4\pi/\kappa A}P_0$ is no longer satisfied and the linear dielectric response is recovered. Things then become quite familiar. Electrons are mainly located between $r_1=\kappa b/Z^*$ and $r_A=\kappa b$ with the majority at radius $r_m=\kappa b/{Z^*}^{1/3}$ as given by the Thomas-Fermi model \cite{Landau}. Although quantum mechanics gives a nonzero electron density at $r<r_1$, the number of total electrons within this radius is only $\sim 1$ and can be ignored. So approximately, when $r_1\gg r^*$, i.e., $\kappa\gg (Z^*)^2\simeq(R/a)^{18/7}$, there's a spatial separation between inner collapsed electrons and outer ones that form the usual Thomas-Fermi atom with the renormalized nucleus. When $\kappa$ is not so big, such separation is absent, which actually happens more often in real situations. The inner tail then connects to the outer electrons with the Thomas-Fermi approach valid all the way and the dielectric response becomes linear at $r=r_c\propto a\kappa^{1/4}{Z^*}^{1/2}$. One should note, as long as $\kappa$ is large enough to satisfy $r_m\gg R$ which gives $\kappa\gg (R/a)^{10/7}$, the majority of the outer electrons located at $r_m$ haven't intruded into the cluster or the highly screening double-layer structure near the cluster surface. The charge renormalization process remains undisturbed and the total net charge seen by outer electrons is still $Z^*$. The corresponding radial electron concentration $n(r)r^2$ is shown in Fig. \ref{fig:3a}. At $\kappa\ll(R/a)^{10/7}$, in most of the space the dielectric response is linear. In that case, almost all electrons reside in the cluster with only some spill-over near the surface. The positive and negative charges are uniformly distributed inside the cluster as described by the Thompson ``jelly" model.

\begin{comment}When there's a periodic system of spherical donor clusters, e.g., a line of charged dots where each dot can approximately be regarded as a spherical donor cluster, at distances much larger than their spatial separation, the system looks like a line of charge or a big cylindrical cluster. To deal with this situation, we address the cylindrical problem in the next section.
\end{comment}

\section{Cylindrical Donor Clusters}

\begin{figure}[h]
\includegraphics[width=0.5\textwidth]{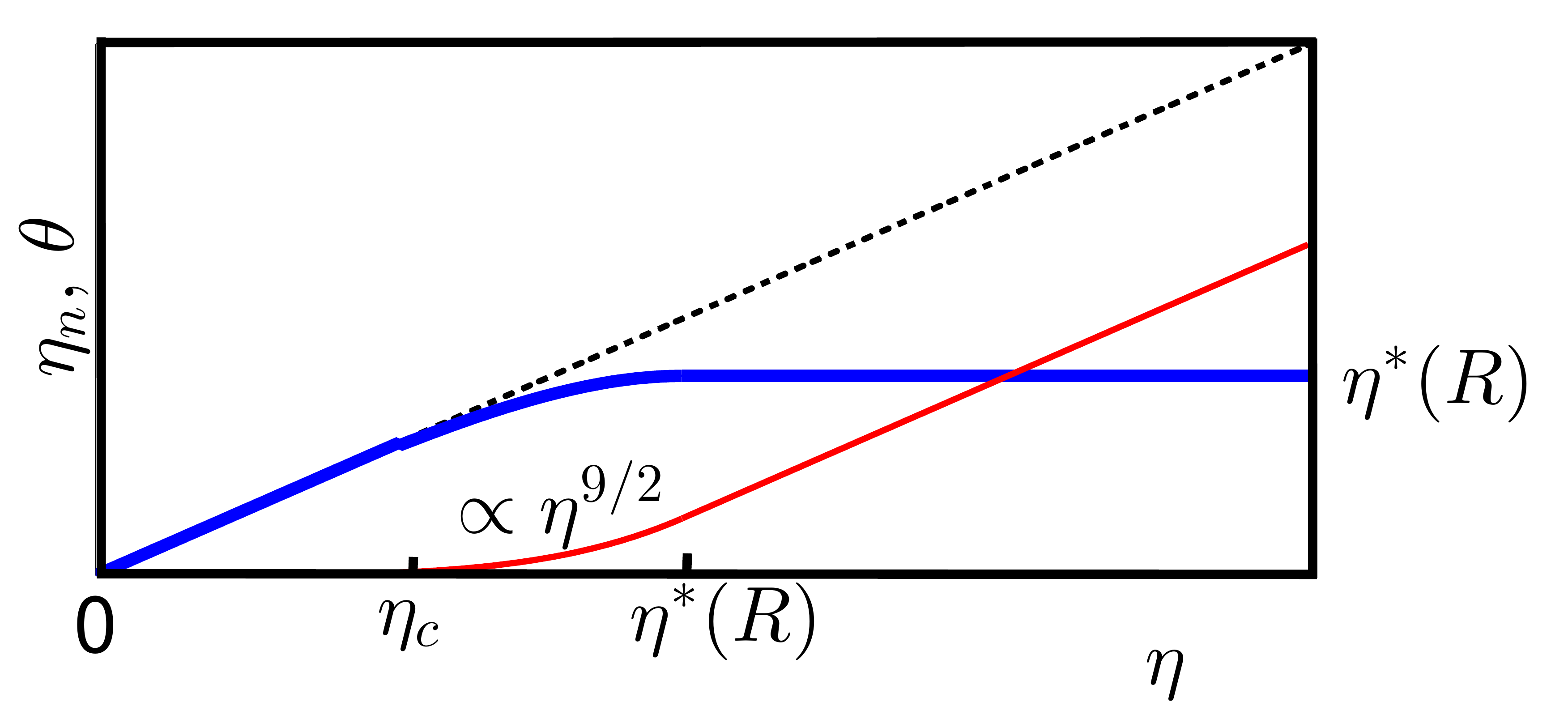}\\
\caption{(Color online) Number of collapsed electrons per unit length $\theta$ and renormalized net charge density $\eta_n$ as a function of cluster charge density $\eta$. The thick solid line (blue) shows $\eta_n(\eta)$. The thin solid line (red) represents $\theta(\eta)$. The dashed line (black) is a guide-to-eye with $\eta_n=\eta$. $\theta(\eta)\propto \eta^{9/2}$ at $\eta_c\ll\eta\ll\eta^*$.}\label{fig:4}
\end{figure}

In some cases, the donor clusters are more like long cylinders than spheres. %It can be that the clusters are created as continuous lines of charge or they're fabricated as a periodic system of separate spheres which at large distances looks similar to a line of charge.
For this situation, a cluster is described by the linear charge density $\eta e$ while its radius is still denoted as $R$. We use a cylindrical coordinate system with the $z$ axis along the axis of the cylinder cluster and $r$ as the distance from the axis.
\begin{comment}
When the dielectric response is linear, close to the cluster, the electric potential is approximately $2\eta e\ln(r/r_0)/\kappa$ where $r_0$ is the radius defined as zero potential position. Since the logarithmic function is slowly varying, the potential is rather flat and the corresponding electron density $n(r)$ is almost constant. The radial electron concentration $n(r)r$ then grows with $r$. At larger distances, one needs to consider the electron screening accounted using the Thomas-Fermi approximation (see Eq. (\ref{eq:5})). It's easy to get the screened potential $\phi(r)\sim (\kappa^2e/b)(b/r)^4$. The radial electron concentration $n(r)r$ then decays as $1/r^5$. Thereby, there's a peak of radial electron concentration at $r\sim r_m=(\kappa b)^{3/4}/\eta^{1/4}$ where the bare cluster potential $\sim \eta e /\kappa$ crosses over to the screened one. Since the majority of electrons are located around $r_m$, at $r\ll r_m$, we can ignore the existence of electrons and the following discussions are based on this approximation.

At small distances from the cluster where $r\ll r_m$, the electric field is big so that the dielectric response is nonlinear. We show below that due to this,\end{comment}
We show that when the charge density $\eta e$ is larger than certain value $\eta_c e$, electrons begin to collapse into the cluster and the charge density is weakly renormalized. When $\eta$ exceeds another value $\eta^*\gg\eta_c$, the renormalization becomes so strong that the net density $\eta_n$ remains $\simeq \eta^*$ regardless of the original density $\eta$ (see Fig. \ref{fig:4}).
Our problem is similar to that of the charged vacuum condensate near superconducting cosmic strings \cite{Vilenkin_1999}, and is also reminiscent of the Onsager-Manning condensation  \cite{Onsager_1967, *Manning_1969} in salty water\footnote{For example, in salty water, the negative linear charge density of DNA is renormalized from $\simeq-4e/l_B$ to the universal net value $-e/l_B$ due to the condensation of Na$^+$ ions onto the DNA surface. Here $l_B=e^2/\kappa_wk_B T\simeq7\,\AA$ where $\kappa_w=81$ is the dielectric constant of water and $T$ is the room temperature.}.

\begin{figure}[h]
$\begin{array}{l}
\includegraphics[width=0.5\textwidth]{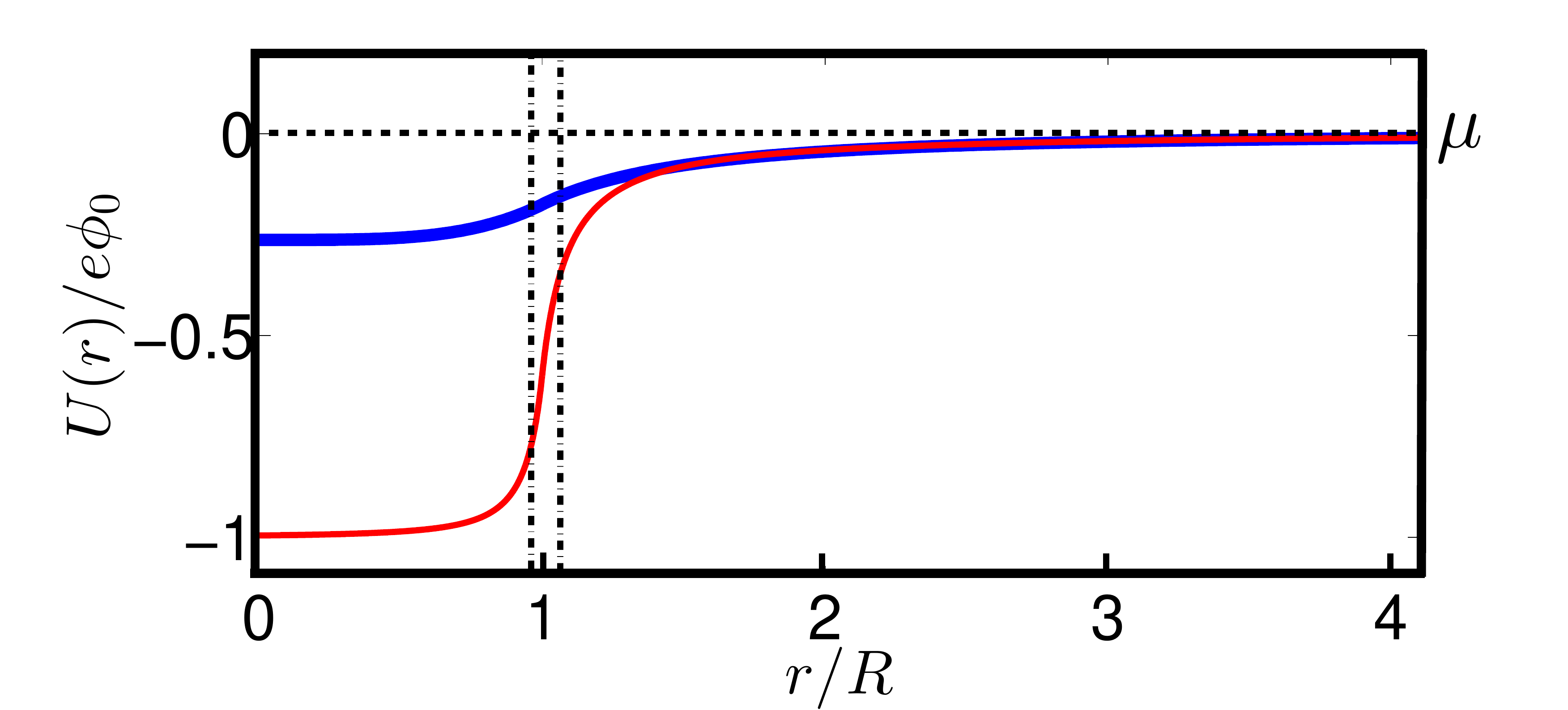}\\
\end{array}$
\caption{(Color online) Potential energy of electrons $U(r)=-e\phi(r)$ as a function of radius $r$. $\phi_0$ is defined as $n(\phi_0)=n_0$, where $n_0=\eta/\pi R^2$, $n(r)$ is a function of $\phi(r)$ given by Eq. (\ref{eq:5}). The thick solid line (blue) represents the potential profile of a cluster of charge density $\eta\lesssim \eta^*$ which is in the regime of weak renormalization of charge. The thin solid line (red) represents the potential of a cluster with $\eta\gg \eta^*$ which is in the strong renormalization regime. The two vertical dotted lines show edges of the ``double-layer" structure of width $\sim d \ll R$. The horizontal dashed line (black) indicates the position of the chemical potential $\mu= 0$.}
\label{fig:5}
\end{figure}

\emph{Renormalization of Linear Charge Density.}---
For a uniformly charged cylindrical cluster with a linear charge density $\eta e$, similar to what we did in Sec. II, we get $D(r)=2\eta(r)e/r$, where $\eta(r)=\eta r^2/R^2$  at $r<R$ and $\eta(r)=\eta$ at $r>R$.  We then can calculate the electric field using Eqs. (\ref{eq:1}) and (\ref{eq:2}) and get the electric potential $\phi(r)$ as:
\begin{subequations}
\begin{align}
\phi(r)&=\frac{A}{P_0^2}\left(\frac{\eta e}{2\pi}\right)^3\left(\frac{3}{4}\frac{1}{R^2}-\frac{1}{4}\frac{r^4}{R^6}\right), 0<r<R\\
\phi(r)&=\frac{A}{P_0^2}\left(\frac{\eta e}{2\pi}\right)^3\frac{1}{2r^2}, \quad\quad\quad\quad\quad\quad\quad R<r\label{eq:14b}
\end{align}
\end{subequations}
with $\phi(r=\infty)$ chosen to be $0$. The corresponding potential energy $U(r)=-e\phi(r)$ is shown in Fig. \ref{fig:5} by the thick solid line (blue). Using the Schrodinger equation and setting the momentum $p\simeq\hbar/2r$, we find that the tightly bound states of electrons, in which electrons are strongly confined within the cluster (at $r<R$), exist only when $\eta>\eta_c$ where
\begin{equation}
\eta_c\approx2\pi\left(\frac{b}{Aa}\right)^{1/3}\frac{1}{a},
\end{equation}
which, contrary to the $Z_c$ value got in the spherical case, does not depend on $R$. Electrons begin to collapse into the cluster at $\eta>\eta_c$ and in the beginning they are located near the axis where the potential energy is lowest (see Fig. \ref{fig:5}). With increasing $\eta$, the electron density grows and one can adopt the Thomas-Fermi description. Using Eq. (\ref{eq:5}) and (\ref{eq:14b}), one gets the electron density $n(r)\propto 1/r^3$ at $r>R$ and the total number of collapsed electron per unit length is
\begin{equation}
\theta=\int_0^\infty n(r)2\pi rdr=0.5\eta\left(\frac{\eta }{\eta^*}\right)^{7/2}\propto\eta^{9/2},
\label{eq:16}
\end{equation}
where
\begin{equation}
\eta^*=\frac{1}{a}\left[2\pi\left(\frac{b}{Aa}\right)^{1/3}\right]^{9/7}\left(\frac{R}{a}\right)^{2/7}.
\label{eq:18}
\end{equation}
%where the coefficient $g_2$ is of order 1.
The net charge density $\eta_ne$ is then renormalized to
\begin{equation}
\eta_n=\eta-\theta=\eta\left[1-0.5\left(\frac{\eta}{\eta^*}\right)^{7/2}\right].
\end{equation}

At $\eta\ll \eta^*$, the renormalization of charge density is weak and $\eta_n$ grows with $\eta$. At $\eta>\eta^*$, the number of collapsed electrons is large and the renormalization effect is strong. Most of the cluster is then neutralized by electrons and the final net density $\eta_n$ is much smaller than $\eta $. Following the logics similar to those in the spherical case, and by using Eq. (\ref{eq:16}), one can show that $\eta_n$ reaches a saturation value of $\eta^*$ at $\eta\gg \eta^*$.% where $\gamma_2$ is a numerical coefficient of order 1.
$ $ The dependence of $\eta_n$ on $\eta$ is shown in Fig. \ref{fig:4}, which resembles Fig. \ref{fig:1}.
\begin{comment}We can connect the results we have got for the cylindrical case with those of spherical clusters. To this end, we look at a long cylinder cluster made of an array of spheres closely aligned with each other. Near each sphere, the effect of others is secondary, and we can approximately use the spherical results which give $Z_c\simeq R/a$ and $Z^*\simeq (R/a)^{9/7}$ for the critical and saturation values of charge. At larger distances, of several $R$, the inhomogeneity of the charge array can be ignored and we can apply the cylindrical results, which give $\eta_c$ independent of $R$ and $\eta^*\simeq (R/a)^{2/7}/a$. Considering that each sphere spreads a width of $\sim R$ along the array axis, the corresponding charge density $\eta e $ it contributes will be $Ze/R$. At $Z=Z_c$ where collapse starts, $\eta=Z_c/R\simeq \eta_c $ has no dependence on $R$. When the renormalized charge saturates at $Z^*$, we get $\eta=Z^*/R\simeq(R/a)^{2/7}/a=\eta^*$. The system transitions from the spherical case to the cylindrical one with all our results matching each other consistently.
\end{comment}

\emph{Radial Distribution of Electrons.}---
\begin{comment}
At $\eta_c<\eta<\eta^*$, the potential decays as $1/r^2$ outside the cluster and the radial concentration of electrons $n(r)r$ diminishes as $1/r^{2}$ as well:
\begin{equation}
n(r)r=c_6\frac{\eta}{a}\left(\frac{\eta}{\eta^*}\right)^{7/2}\left(\frac{R}{a}\right)\left(\frac{a}{r}\right)^2
\label{eq:19}
\end{equation}
where $c_6\approx 0.05$.
\end{comment}
At $\eta\gg \eta^*$, there are lots of collapsed electrons inside the cluster where $n(r)=n_0=\eta/\pi R^2$ and the potential energy is low. Again, there's a ``double-layer" structure on the surface that provides steep growth of potential energy with $r $ at $r=R$.
\begin{comment}
The Thomas-Fermi approach gives the differential equation of $\phi(r)$ in the cylindrical coordinates:
\begin{subequations}
\begin{align}
\left(\frac{d}{dr}+\frac{1}{r}\right)\left(\frac{d\phi}{dr}\right)^{1/3}&=\frac{A^{1/3}e}{P_0^{2/3}}(n(r)-n_0),\, r<R\label{eq:21a}\\
\left(\frac{d}{dr}+\frac{1}{r}\right)\left(\frac{d\phi}{dr}\right)^{1/3}&=\frac{A^{1/3}e}{P_0^{2/3}}n(r), \quad\quad\,\quad r>R\label{eq:21b}
\end{align}
\end{subequations}
\end{comment}
Close to the cylinder surface at $0<r-R\ll R$, as for the sphere, we can approximately use a plane solution of $\phi(r)$ as given by Eq. (\ref{eq:10}). The expression of the characteristic decay length $d$ is also the same as in Eq. (\ref{eq:13}).
\begin{comment}
\begin{equation}
\begin{aligned}
n(r)r&=\frac{c_4}{A^{3/7}}\frac{1}{b^3}\left(\frac{b}{a}\right)^{24/7}\left(\frac{a}{x+d}\right)^{12/7}r\\
&\approx\frac{c_4}{A^{3/7}}\frac{1}{b^3}\left(\frac{b}{a}\right)^{24/7}\left(\frac{a}{x+d}\right)^{12/7}R,
\end{aligned}
\label{eq:20}
\end{equation}
where $x=r-R\ll R$. \end{comment}
When $x=r-R$ is comparable to $R$, the plane solution crosses over to the fast decaying potential $\propto 1/r^2$ as given by Eq. (\ref{eq:14b}) with $\eta$ replaced by $\eta_n\simeq\eta^*$. %The corresponding radial electron density $n(r)r$ is expressed by Eq. (\ref{eq:19}) also with $\eta$ replaced by $\eta_n$.
A schematic plot of the potential energy $U(r)=-e\phi(r)$ is shown in Fig. \ref{fig:5}.% while the corresponding radial distribution of electrons $n(r)r$ is shown in Fig. (\ref{fig:6})

This potential produces a universal tail of electron density $n(r)\sim 1/r^3$. The corresponding radial electron concentration $n(r)r$ is $\sim 1/r^2$. Since the Fermi wavelength $\lambda\simeq n(r)^{-1/3}$, we get $\lambda\sim r$, i.e., the Thomas-Fermi approach is only marginally valid. The collapsed electrons extend until the linear dielectric response is recovered and then connect to the outer electrons.

\section{Finite-temperature ionization of cluster atoms and its experimental implications.}

So far, we dealt with very low temperature. At a finite temperature $T$, the neutral cluster atom can get ionized due to the entropy gain of ionized electrons.
The donor cluster atom becomes a positive ion with charge $Z_i(T)e$. Our goal below is to find this charge.

We assume that we have a small but finite three-dimensional concentration $N$ of spherical clusters and the charge $Z_i(T) < Z^*$, i. e., the outer electron shell is still incompletely ionized. Such a cluster can bind electrons with an ionization energy $Z_i(T)^{2}e^2/\kappa^2 b$. We can find $Z_i(T)$ by equating this energy with the decrease in the free energy per electron $k_BT\ln(n_0/n)$ due to the entropy increase, where $k_B$ is the Boltzmann constant,
$n=Z_i(T)N$ is the concentration of ionized electrons and $n_0=2/\lambda^3$ with $\lambda=\sqrt{2\pi\hbar^2/m^*k_BT}$ as the DeBroglie wavelength of free electrons at temperature $T$. At $\kappa=20000$, $b=0.29\,\AA$, $m^*=1.8m_e$ where $m_e$ is the electron mass and $N=10^{15}$ cm$^{-3}$ (estimated from that the concentration of total donor electrons is around $10^{18}$ cm$^{-3}$ and each cluster contributes $\sim 300$ donor electrons), we get $Z_i(T)\gtrsim Z^*$ at $T\gtrsim 8$ K with $Z^*=100$ which is a reasonable estimate. This shows that the outer electrons are completely ionized at temperatures that are not too low. For the inner core electrons, the dielectric response is nonlinear and the attractive potential is stronger. So the ionization energy is higher $\simeq A(Z^*e/4\pi)^3/5P_0^2R^5$ for electrons at $r\simeq R$. At $R=4a$, it is found that only at $T>450$ K can
the inner electrons be ionized by a considerable quantity (the $1/r^{15/2}$ tail is completely stripped then). So the inner electrons are robust against the thermal ionization.

For the cylindrical cluster, since it can effectively be regarded as the assembly of sphere clusters, one can expect that electrons are harder to be stripped off than in the spherical case. The thermal ionization is thus somewhat weaker. For the outer electrons, the dielectric response is linear and the ionization degree is determined by the Onsager-Manning linear density $\eta_{OM}$ \cite{Onsager_1967,*Manning_1969}. It can be derived as follows. The potential energy of electrons caused by the cylindrical charge source with a linear density $\eta_i e$ grows with the radius $r$ as $2\eta_{i}e^2\ln(r/r_0)/\kappa$ while  the entropy increases as $2k_B\ln(r/r_0)$, where $r_0$ is a chosen reference point. In equilibrium, by equating the energy increase and the entropic decrease of the free energy one gets the critical concentration $\eta_i=\eta_{OM}=\kappa k_BT/e^2$. This is the universal value of the net charge density, which depends only on the temperature $T$ and the dielectric constant $\kappa$ of the media similar to the case of DNA in salty water\footnotemark[1].

When the outer electrons are completely ionized, the charge density is expected to be $\eta^*e$ given by Eq. (\ref{eq:18}). Taking $R=4a,\,A=0.9,\,\kappa=20000,\,a=3.9\,\AA$ and $b=0.29\,\AA$, we get $\eta_{OM}\gtrsim\eta^*$ at $T\gtrsim 10$ K.
Thus, the ionization of the outer shell proceeds until $T$ grows to 10 K.
Since the product $\kappa T$ is almost fixed at $T > 10$ K, $\eta_{OM}$ actually stops growing with the temperature at $T > 10 $ K. This practically means that only outer electrons are ionized at  $T > 10 K$. The inner electrons are well preserved against the ionization even at  $T > 10$ K.

Thus, in both spherical and cylindrical cases, the outer electrons are thermally ionized at not very low temperature while the inner ones are mainly kept by the cluster. The final observable charge or charge density is then equal to $Z^*$ or $\eta^*$. Below we will discuss how one can observe the thermal ionization.

Experimentally, charged clusters can be created controllably on the surface of LAO/STO structure when the LAO layer is of subcritical thickness $\lesssim 3$ unit cell \cite{Cen_2008, *Cen_2010}. A conducting atomic force microscope (AFM) tip is placed in contact with the top LaAlO$_3$ (LAO) surface and biased at certain voltage with respect to the interface, which is held at electric ground. When the voltage is positive, a locally metallic interface is produced between LAO and STO where some positive charges are accumulated in the shape of a disc. The same writing process can also create a periodic array of charged discs.

Let's first concentrate on a disc of positive charge created in this manner on the STO surface. Close to the surface and in the bulk STO, one should apply the plane solution given by Ref. \onlinecite{RS} and repeated by Eq. (\ref{eq:10}) above. When the distance $r$ from the disc center is large, i.e., $r\gg R$, the disc behaves like a charged sphere. Our results for a sphere are still qualitatively correct in this case.

In a periodic array of highly charged discs with period $2L$, the linear concentration of free electrons responsible for the conductance at a very low temperature is of the order of $n(L)L^2$, where $n(r)$ is the electron density around a spherical donor cluster given by Sec. II. When the overlapping parts between neighboring discs belong to the outer electron shells, the corresponding density at $r=L$ is that of a Thomas-Fermi atom with charge $Z^*$. In this situation, the overlapping external atmosphere forms conductive ``bridges" between discs at low temperature. When $T$ increases, however, the outer electrons are ionized and the bridges are gone. These free electrons spread out over the bulk STO. At $T\lesssim 30$ K, electrons are scattered mainly by the Coulomb potential of donors and the corresponding mobility decreases with a decreased electron velocity. For the electrons ionized into the vast region of the bulk STO, they're no longer degenerate, so their velocity becomes much smaller at relatively low temperature. This results in a much smaller mobility of the ionized electrons than those bound along the chain. Their contribution to the conductivity is thus negligible. The system becomes more resistive due to the ionization and one can observe a sharp decrease of the conductivity along the chain. It may be possible for the chain to transition from a conducting ``line" to multi-quantum-dots. Interesting phenomena of the conductance behavior such as the Coulomb blockade can probably emerge.

\section{Conclusion}

In this paper we have studied the structure of a many-electron ``atom" whose center is a strongly charged donor cluster. It is determined by the
collapse of electrons to the cluster in SrTiO$_3$ due to the nonlinear dielectric response at small distances from the cluster surface. For a spherical cluster, when its charge number exceeds a critical value $Z_c$, the potential well inside it becomes deep enough to trap electrons despite their high kinetic energy. In the beginning, the cluster charge $Ze$ is weakly renormalized by the electrons. When $Ze$ grows beyond another value $Z^*$, the number of collapsed electrons becomes large so that the most of the cluster is neutralized and the charge is strongly renormalized to $Z^*$. This strong renormalization is realized via a ``double-layer" structure on the cluster surface. The corresponding potential profiles and radial distributions of electrons are investigated. The critical and saturation values are found to be dependent only on the cluster radius: $Z_c\simeq(R/a)$ and $Z^*\simeq(R/a)^{9/7}$. At zero temperature, a renormalized cluster with charge $Z^*$ is the nucleus of a Thomas-Fermi atom. This nucleus is surrounded by the external electron atmosphere which is sparse due to the weak Coulomb interaction at a large dielectric constant. These outer electrons can easily be stripped off by the thermal ionization leaving only the compact ionic core with charge $Z^*$. The case of a cylindrical donor cluster is discussed as well where similar results are found. Namely, when its linear charge density $\eta e$ is larger than $\eta_c e\simeq e/a $, electrons start collapsing to the cluster. At $\eta \gg \eta^*$
where $\eta^* \simeq  (R/a)^{2/7}/a $ is the saturation density, the net charge density of the cluster $\eta_ne$ stays at the level of $\eta^*e$.  At zero temperature, this renormalized cylindrical cluster is surrounded by a sparse electron atmosphere which can be ionized at temperatures larger than 10 K.
We also discuss how one can verify our predictions by measuring the conductivity of a chain of charged discs on the surface of LAO/STO structures.
$\phantom{}$
\vspace*{2ex} \par \noindent
{\em Acknowledgments.}

We are grateful to B. Skinner, E. B. Kolomeisky, B. Jalan, C. Leighton, A. P. Levanyuk and J. Levy for careful reading of the manuscript and valuable advice, and to A. Kamenev, D. L. Maslov, A. Vainshtein and M. B. Voloshin for helpful discussions. This work was supported primarily by the National Science Foundation through
the University of Minnesota MRSEC under Award No. DMR-1420013.
%\end{small}

\bibliography{TF}

\end{document}